\documentstyle[11pt,newpasp,psfig]{article}

\def\psfig#1{\relax}
\def\gtsim{\mathrel{\hbox{\rlap{\lower.55ex \hbox {$\sim$}}
                   \kern-.33em \raise.5ex \hbox{$>$}}}}
\def\ltsim{\mathrel{\hbox{\rlap{\lower.55ex \hbox {$\sim$}}
                   \kern-.27em \raise.45ex \hbox{$<$}}}}
\def\etal{{\it et al.}}
\def\eg{{\it e.g.}}

\begin{document}
\title{Most Real Bars are Not Made by the Bar Instability}
\author{J. A. Sellwood}
\affil{Department of Physics \& Astronomy, Rutgers University, \\
136 Frelinghuysen Road, Piscataway NJ 08854-8016, USA \\
e-mail: sellwood@astro.rutgers.edu}

\begin{abstract}
Having once thought that we understood how some galaxies were barred and had 
difficulty accounting for the absence of bars in others, it now seems that we 
have the opposite problem.  Most real galaxies have centres dense enough to 
inhibit bar formation, even if they have massive discs.  This is also true 
for barred galaxies, which therefore could not have acquired their bars as a 
result of a self-excited, global instability.  There is a hint from the 
morphology of galaxies in the Hubble Deep Fields that the growth of bars 
might be a slow or secular process.  Here I discuss possible mechanisms that 
could form bars long after the disc is assembled.
\end{abstract}

\section{Introduction}
I am probably not the only person who once thought that the bar instability 
was responsible for the bars we see in a decent fraction of nearby galaxies 
(Sellwood \& Wilkinson 1993).  While understanding the {\it absence\/} of 
strong bars in most galaxies was a major headache for galaxy dynamicists 
(Ostriker \& Peebles 1973; Toomre 1974), we took comfort from the fact that 
we thought we knew why some galaxies have bars.  Now we understand why 
galaxies are not barred, we can no longer claim to comprehend the origin of 
bars -- our problem has inverted!  

I have two reasons for doubting that bars, especially those in bright 
galaxies, could have formed through the usual global instability.  The first, 
which has been evident for some time, is that many barred galaxies have 
strong inner Lindblad resonances (ILRs) and the second is the recent claim by 
Abraham \etal\ (1999; see also Merrifield, this volume) that bars were less 
common at $z>0.5$.  The latter result is clearly still quite tentative but, 
even if it went away, the ILRs in bars are themselves ample evidence against 
straightforward instabilities.

\section{Inner Lindblad resonances in bars}
A number of lines of evidence all suggest that many bars possess ILRs.  
Recall that the gravitational stresses from a bar drive gas inwards towards 
the center until an ILR is encountered, where it piles up in a ring (\eg\ 
Athanassoula 1992).  Optical nuclear rings with diameters of a few hundred pc 
(Buta \& Crocker 1993), often the sites of vigorous star formation, are seen 
in many, though not quite all, barred galaxies; a particularly beautiful case 
is the HST image of NGC~4314 (Benedict \etal\ 1998).  Ring-like 
concentrations of molecular gas, often with ``twin peaks'' morphology, are 
now being observed with molecular interferometer arrays (\eg\ Helfer \& Blitz 
1995; Turner 1996; Kenney 1997; Sakamoto \etal\ 1999).  Further, Athanassoula 
(1992) argued that gas flow models could produce shocks at the positions of 
the offset dust lanes along the bar in many galaxies only if a strong ILR 
were present.  Those barred galaxies for which the observed gas velocity 
field has been modelled all appear to have ILRs (Duval \& Athanassoula 1983; 
Lindblad \etal\ 1996; Regan \etal\ 1997; Weiner \etal\ 1999).  Finally, there 
is strong evidence for an ILR in the bar of the Milky Way (Binney \etal\ 
1991; Weiner \& Sellwood 1999).

\subsection{Disc Stability}
Much of this overwhelming body of evidence in favour of a central density 
high enough to ensure an ILR has been known for some time.  Yet it did not 
seem to represent much more than a nagging worry because we did not fully 
understand how galaxy disks were stabilized.  Now that we know a high central 
density really does stabilize a galaxy, this minor worry has suddenly become 
serious.

Toomre (1981) argued that a dense centre could prevent the bar instability by 
inserting an ILR to cut the feedback to the swing-amplifier.  Only numerical 
simulations with reasonable particle numbers and good time and spatial 
resolution are able to reproduce the correct behaviour in the central regions 
and confirm this prediction.  They have now established that galaxy models 
containing massive discs can be dynamically cool and yet not form bars 
(Sellwood 1985; Sellwood \& Moore 1999; Sellwood 1999).  Rubin \etal\ (1997) 
and Sofue \etal\ (1999) show that virtually {\it all\/} bright galaxies 
($V_{\rm max} \gtsim 150\hbox{ km s}^{-1}$) have dense centres -- the reason 
for the stability of real galaxies is now clear.

If the mass distribution in barred galaxies today is such that it should have 
inhibited a bar from forming, why are these galaxies barred?  We can dismiss 
two obvious ideas.  If bars formed with much higher pattern speeds and have 
since slowed down (without getting longer, see below) then co-rotation would 
lie well beyond the end of the bar, which contradicts much of the evidence 
already cited as well as direct measurements (Merrifiend \& Kuijken 1993; 
Gerssen \etal\ 1999).  Perhaps the mass distribution was originally more 
uniform, but enough gas has subsequently been driven into the centre to 
create the ILR.  This idea seems physically reasonable since as little as 
1--2\% of the galaxy mass, together with the supporting response of the 
stars, is sufficient (Sellwood \& Moore 1999).  However, this same process 
weakens or destroys the bar, as has been argued by Norman and his co-workers 
(Hasan \& Norman 1990; Pfenniger \& Norman 1990) and reproduced in 
simulations (Friedli 1994; Norman \etal\ 1996).

\section{Hubble Deep Fields}
The study of the barred galaxy fraction as a function of redshift by Abraham 
\etal\ (1999; see also Merrifield, this volume) raises a further difficulty 
for the bar instability picture.  They find very few strongly barred galaxies 
at $z>0.5$, suggesting that bars develop long after the discs of these 
galaxies are assembled.  This result may suggest a gradual build-up of 
the disc until the rapid dynamical instability occurs.  However, late 
infalling material probably contributes to the outer disc, and not to the 
central density (\eg\ Simard \etal\ 1999), and so will have less effect on 
global stability and, furthermore, such an idea would not avoid the 
stabilizing effect of the observed dense centres.

\section{Other Bar-Formation Mechanisms}
The above discussion suggests that we should abandon the idea that bars are 
caused by the global dynamical instability.  If this most obvious mechanism 
for bar formation is excluded, what are the alternatives?

One possibility is an encounter with another galaxy which triggers a bar 
(\eg\ Noguchi 1987; Gerin \etal\ 1990; Mihos \etal\ 1997).  There is some 
evidence for higher barred fraction in dense environments (Elmegreen \etal\ 
1990; Giuricin \etal\ 1993), suggesting that this does occur in practice.  
However, the idea is unattractive for two reasons: first, interactions were 
more common in the early universe, so the bar fraction should build up 
quickly, in contradiction to Abraham \etal\ ~Second, Miwa \& Noguchi (1998) 
find that bars formed through tidal encounters generally have rather low 
pattern speeds, whereas most bars are believed to rotate rapidly, as noted 
above.

Lynden-Bell (1979) argued for a gradual secular bar growth through orbit 
trapping.  However, his mechanism would again form bars having slow figure 
rotation, whereas almost all evidence points to rapid figure rotation.

I currently favour episodic growth, which I reported in some of my early 
simulations (Sellwood 1981).  In this process, a short, weak bar can become 
longer and stronger through trapping of erstwhile disc particles into the 
bar;  strong spiral patterns, which carry away angular momentum, can add many 
particles to  the bar.  It differs from Lynden-Bell's mechanism because 
changes occur in $\sim 1$ orbital period and depend crucially on the phase of 
the spiral relative to the bar.  After one such spiral pattern, the bar is 
significantly longer and slightly slower than before, but co-rotation remains 
just beyond the end of the bar.  It should be noted, hovever, that all 
simulations so far in which I have witnessed this process have required an 
initial seed bar.

\section{Conclusions}
If bars were formed by the global bar instability, then (1) they probably 
should form a bar early in a galaxy's life and (2) they should not form when 
the centre is dense.  Both predictions are inconsistent with observations, 
the second much more decisively, arguing strongly that bars were not formed 
in this manner.

Thus an alternative bar-forming mechanism is needed.  I propose one such 
possibility, but the idea is not fully worked out.  Ideally some 
observational test is needed that would be able to distinguish a bar formed 
through this, or any other secular process, from one formed through a global 
dynamical instability.  It would also be desirable to be able to predict the 
distribution of bar strengths in galaxies today, although this may have to 
await substantial progress in our understanding of the late stages of galaxy 
formation.

\acknowledgments
This work was supported by NSF grant AST 96/17088 and NASA LTSA grant NAG 
5-6037.

\end{document}